\documentclass[12pt,a4paper]{article}
\usepackage{amssymb,amsmath,amsthm}
\usepackage{bbm,latexsym}
\usepackage{pifont}

\newcommand{\ord}{\mbox{ord}\, G}
\newcommand{\mnu}{\mathcal{M}_\nu}
\newcommand{\dsm}{\Delta m^2_\odot}
\newcommand{\dam}{\Delta m^2_\mathrm{atm}}

\newcommand{\para}{\paragraph}
\newcommand{\oo}{\mbox{ord}\,}
\newcommand{\zz}{\mathbbm{Z}_2}
\newcommand{\one}{\mathbf{1}}
\newcommand{\two}{\mathbf{2}}
\newcommand{\three}{\mathbf{3}}

\newtheorem{theorem}{Theorem}

\allowdisplaybreaks

\begin{document}
\begin{center}
{\bfseries 
Theory of Neutrino Masses and Mixing\footnote{\small Lecture presented at 
IV International Pontecorvo
Neutrino Physics School, 
September 26 -- October 6, 2010, Alushta, Crimea}}
\vskip 5mm
Walter Grimus
\vskip 5mm
{\small\it University of Vienna, Faculty of Physics, Boltzmanngasse 5, \\ 
           A--1090 Vienna, Austria}
\end{center}
\vspace{5mm}
\centerline{\bf Abstract}
We motivate the usage of finite groups as symmetries of the Lagrangian. After
a presentation of basic group-theoretical concepts, we introduce the notion of
characters and 
character tables in the context of irreducible representations and discuss
their applications. We exemplify these theoretical concepts with the
groups $S_4$ and $A_4$. Finally, we discuss the relation between tensor
products of irreducible representations and Yukawa couplings and describe a
model for tri-bimaximal lepton mixing based on $A_4$.
\vskip 10mm

\section{Introduction}
\paragraph{Motivation for horizontal symmetries:} 
The mass spectrum of quarks and leptons is one of the least understood
facts of particle physics. However, it was noticed quite
early~\cite{gatto} that the Cabbibo angle might be a function of the
ratio of down and strange quark mass because numerically one has
\begin{equation} \label{c}
\sin\theta_c \simeq \sqrt{\frac{m_d}{m_s}}.
\end{equation}
A very popular possibility to generate fermion masses and mixing is
the Higgs mechanism. This has brought about the idea that in such a
framework the CKM matrix could be explained by symmetries 
acting on the three quark families which restrict the Yukawa
couplings such that a relation like equation~(\ref{c}) becomes
possible. Since the CKM matrix is not far from the unit matrix and the
up and down quark mass spectra are strongly hierarchical, is seems at
least plausible that the mixing angles are functions of quark mass
ratios. 

The observation by Harrison, Perkins and Scott~\cite{HPS} that lepton
mixing is in good approximation \emph{tri-bimaximal}, i.e. compatible
with the mixing matrix 
\begin{equation}\label{HPS}
U \simeq \left( \begin{array}{rrr}
2/\sqrt{6} & 1/\sqrt{3} & 0 \\ 
-1/\sqrt{6} & 1/\sqrt{3} & -1/\sqrt{2} \\ 
-1/\sqrt{6} & 1/\sqrt{3} & 1/\sqrt{2}
\end{array} \right) \equiv U_\mathrm{HPS},
\end{equation}
has given a boost to the idea of family symmetries. 
In the lepton sector it seems that mixing angles could be related to 
``pure numbers.'' At any rate, $U$ is very different from the unit matrix
and thus lepton mixing is very different from quark mixing~\cite{results}.

\paragraph{Neutrino mass spectrum:} The idea that the elements of $U$ are, in
good approximation, pure numbers (and not functions of lepton mass ratios)  is
in accord with the observation of the neutrino 
mass spectrum: it is either completely different from the charged-fermion mass
spectra or its hierarchy is not so pronounced~\cite{results}. 

We know from neutrino oscillations that the neutrino mass spectrum is 
non-degenerate. 
\begin{figure}
\begin{center}
\setlength{\unitlength}{1cm}
\begin{picture}(4,4)
\put(1,1){\vector(0,1){3}}
\put(3,1){\vector(0,1){3}}
\thicklines
\put(0.7,1.3){\line(1,0){0.6}}
\put(0.7,1.5){\line(1,0){0.6}}
\put(0.7,3.5){\line(1,0){0.6}}
\put(1.1,1.1){\makebox(0,0)[l]{$m_1$}}
\put(1.1,3.65){\makebox(0,0)[l]{$m_3$}}
\put(1.1,1.65){\makebox(0,0)[l]{$m_2$}}
\put(1,0.8){\makebox(0,0)[t]{\tt normal}}
\put(0.6,1.4){\makebox(0,0)[r]{$\Delta m^2_\odot$}}
\put(0.8,2.5){\makebox(0,0)[r]{$\Delta m^2_\mathrm{atm}$}}
\put(2.7,1.3){\line(1,0){0.6}}
\put(2.7,3.3){\line(1,0){0.6}}
\put(2.7,3.5){\line(1,0){0.6}}
\put(3.1,1.1){\makebox(0,0)[l]{$m_3$}}
\put(3.1,3.65){\makebox(0,0)[l]{$m_2$}}
\put(3.1,3.1){\makebox(0,0)[l]{$m_1$}}
\put(3,0.8){\makebox(0,0)[t]{\tt inverted}}
\put(3.7,3.4){\makebox(0,0)[l]{$\Delta m^2_\odot$}}
\put(3.7,2.3){\makebox(0,0)[l]{$\Delta m^2_\mathrm{atm}$}}
\put(2,0.4){\makebox(0,0)[t]{\tt spectrum}}
\end{picture}
\end{center}
\caption{Types of neutrino mass spectra \label{nuspec}}
\end{figure}
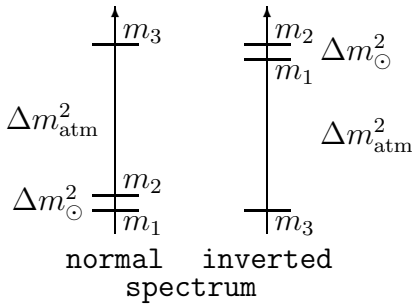
The neutrino mass spectrum is called 
hierarchical, if $m_1 \ll \dsm$, where $m_1$ is
the smallest neutrino mass and $\dsm$ the solar mass-squared difference. Since 
$\dam/\dsm \sim 30$ we conclude that 
$m_3/m_2 \simeq \sqrt{\dam/\dsm} \sim 5 \div 6$ in the hierarchical
case which illustrates that a neutrino mass hierarchy can only be rather
weak. The quantity $\dam$ is the atmospheric mass-squared difference. 
An inverted hierarchy is also possible if (by the usual convention) $m_3$ is
the smallest mass with $m_3 \ll \dam$. Experimentally, the question of the
neutrino mass spectrum is completely undecided. If the 
smallest neutrino mass is denoted by $m_s$ we have a  
{\tt normal} ordering for $m_s = m_1$ and  
an {\tt inverted} ordering for $m_s = m_3$. 
The spectrum is called quasi-degenerate if $m_1 \simeq m_2 \simeq m_3$. 
Of course, also a spectrum between hierarchical and quasi-degenerate is
allowed for both orderings.

\paragraph{Neutrino mass terms and parameter counting:} 
In the following we assume that 
\begin{itemize}
\item[\ding{114}]
neutrinos have Majorana nature and 
\item[\ding{114}]
the charged-lepton mass matrix is diagonal.
\end{itemize}
Majorana neutrinos are theoretically more appealing than Dirac neutrinos
because many mechanisms for neutrino mass generation, e.g. the seesaw
mechanism~\cite{seesaw}, naturally lead to Majorana nature.
The second assumption is used only for the time being for the purpose of
parameter counting. 

A Majorana neutrino mass term is given by
\begin{equation}
\mathcal{L}_\mathrm{Maj} = \frac{1}{2} \nu_L^T C^{-1} \mnu \nu_L + 
\mathrm{H.c.}
\end{equation}
with the charge-conjugation matrix $C$. 
From the anticommutation property of the neutrino fields we conclude that
$\mnu = \mnu^T$, i.e. $\mnu$ is a symmetric but in general complex matrix. For
the transformation to the mass eigenfields, the following theorem, specialized
to $3 \times 3$ matrices, is applied. 
\begin{theorem}[Schur] \label{schur}
For every complex, symmetric matrix $\mnu$ there exists a unitary matrix $U$
with $U^T \mnu U = \mbox{diag}\, (m_1, m_2, m_3)$ and $m_j \geq 0$.
\end{theorem}
\noindent
The matrix $U$ diagonalizing the neutrino mass matrix $\mnu$ is called the 
Pontecorvo--Maki--Nakagawa--Sakata (PMNS) or lepton mixing matrix $U$, provided
we are in a basis where the charged-lepton mass matrix is diagonal. 
The matrix $U$ is usually parameterized as 
\begin{equation}
U = 
e^{i\hat \alpha}
U_{23} U_{13} U_{12}\,
{\rm diag} \left( 1, e^{i \beta_2}, e^{i \beta_3} \right).
\end{equation}
The diagonal phase matrix 
$e^{i\hat \alpha} = \mathrm{diag}\,
\left( e^{i \alpha_1}, e^{i \alpha_2}, e^{i \alpha_3} \right)$ is unphysical
in the charged-current interaction because it can be absorbed into the
charged lepton fields. The matrices $U_{23}$, $U_{13}$ and $U_{12}$ are
rotations in the subsectors indicated by their subscripts:
\begin{eqnarray}
U_{23} &=& \left( \begin{array}{ccc} 1 & 0 & 0 \\
0 & c_{23} & s_{23} \\ 
0 & - s_{23} & c_{23} \end{array} \right), \\
U_{13} &=& 
\left( \begin{array}{ccc} c_{13} & 0 & s_{13} e^{-i \delta} \\
0 & 1 & 0 \\ - s_{13} e^{i\delta} & 0 & c_{13} \end{array} \right), \\
U_{12} &=&
\left( \begin{array}{ccc} c_{12} & s_{12} & 0 \\
- s_{12} & c_{12} & 0 \\ 0 & 0 & 1 \end{array} \right).
\end{eqnarray}
In the mixing matrix the conventions  
$0^\circ \leq \theta_{ij} \leq 90^\circ$ are imposed. As a consequence, one
must allow the full range $0^\circ \leq \delta < 360^\circ$ of the
CP-violating CKM-type phase $\delta$. As for the neutrino masses, one imposes 
$m_1 < m_2$ with $\dsm = m_2^2 - m_1^2$. With this convention the sign of 
$m_3^2 - m_1^2$ is a physical quantity and must eventually be determined by
experiment. 

In summary there are nine physical parameters in neutrino masses and mixing:
three masses, three angles and the three phases $\delta$, $\beta_2$ and
$\beta_3$. The latter two phases are the so-called Majorana phases; if
neutrinos have Dirac nature, they can be removed from the charged-current
interactions by absorbing them into the neutrino fields. 

Let us compare the number of nine parameters with the number of parameters 
in $\mnu$. There are 
$6 \times 2 = 12$ real parameters in $\mnu$. However, e.g. the 
first line and first column can be made real by a phase transformation 
$e^{i\hat \alpha}$ which has no effect in the charged current
interactions---see above. Thus we have nine real physical parameters in $\mnu$
corresponding to the nine physical quantities above. 
As mentioned before, there is also one discrete physical parameter, 
namely $\mbox{sign}\,(m_3^2 - m_1^2)$, which is $+1$ for
the normal ordering and $-1$ for the inverted ordering of the neutrino mass
spectrum. 

Finally, we want to make some remarks concerning the diagonalization of $\mnu$
with theorem~\ref{schur}. If we write $U = (u_1, u_2, u_3)$ with an
orthonormal (ON) basis $u_j$ of $\mathbbm{C}^3$, 
theorem~\ref{schur} tells us that
\begin{equation}
\mnu u_j = m_j u_j^*.
\end{equation}
Note the following points: 
\begin{itemize}
\item
In general, $u_j$ is not an eigenvector of $\mnu$, 
this is the case only for real $u_j$. 
\item
If $\lambda$ is an eigenvalue of $\mnu$, then $|\lambda|$ is in general
\emph{not} a neutrino mass. 
\item
However, the neutrino masses can be obtained by 
$\mnu^\dagger \mnu u_j = m_j^2 u_j$.
\end{itemize}

In these lecture notes we will discuss some features of model building for
lepton masses and mixing---see for instance~\cite{grimus,review} for
reviews. However, we will first delve into useful theoretical aspects of
finite groups and review two groups popular in model building. 
For the general theory of groups we refer the reader
e.g. to~\cite{hamermesh,ramond}. Recent reviews on finite subgroups of $SU(2)$
and $SU(3)$ are presented in~\cite{ludl,tanimoto}, for more specialized recent
reviews see~\cite{gl-2010,ludl-u3,parattu}.

\section{Theory of finite groups}
\subsection{Basics}
\label{basics}
We assume familiarity of the reader with the very basic notions like the
definition of a group, representation, 
irreducible representation (irrep), subgroup, coset and normal subgroup,
which can be found in any text book on group theory, 
e.g. in~\cite{hamermesh,ramond}.

Now we will explain some basic useful concepts. 
On a group $G$ one always has an equivalence relation via the following 
definition: $g_1$  is \emph{conjugate} to $g_2$ if it exists a $g \in G$
such that $g g_1 g^{-1} = g_2$. The sets of equivalent elements are called 
\emph{conjugacy classes}. Obviously, $\{ e \}$ is a class consisting only of
the unit element, and 
a normal subgroup consists of complete conjugacy classes.

\para{Irreps and proper normal subgroups:}
Using a symmetry group in physics mostly boils down to applying its 
irreps to physical objects (multiplets). Therefore, we need to know the
irreps or methods how to track them down. A good part of this section is
devoted to this subject. 

Knowing the proper normal subgroups of $G$ helps in this respect. The notion
``proper'' means that the subgroup is larger than $\{ e \}$ and smaller than
$G$. 
Let $H$ be a proper normal subgroup of $G$, then 
\begin{itemize}
\item 
the mapping $f:\,g \in G \to Hg \in G/H$ is a
homomorphism, i.e. the relation $f(g)f(g') = f(gg')$ holds 
$\forall\,g,g' \in G$, 
\item
and any representation $D$ of $G/H$ induces naturally a representation $\bar D$
of $G$ via $\bar D(g) \equiv D(Hg)$.
\end{itemize}

\para{Direct product:} With two groups $G$ and $G'$ one can form the direct
pro\-duct group $G \times G'$ with the multiplication law 
$(g_1,g_1')(g_2,g_2') = (g_1 g_2,g_1' g'_2)$. This is often used in model
building.  
E.g., one has a symmetry group like the permutation group $S_3$ and enlarges
it by a sign transformation leading to $S_3 \times \mathbbm{Z}_2$,
a direct product of $S_3$ with the cyclic group $\mathbbm{Z}_2$.

\para{Semidirect product:} This generalization of the direct product is
written as $H \rtimes_\phi G$, which symbolizes that 
$G$ acts on $H$ via the homomorphism $\phi:\, G \to \mbox{Aut}(H)$ where
$\mbox{Aut}(H)$ is the group of automorphisms on $H$.
(An automorphism $\phi$ on $H$ is simply a group isomorphism 
$\phi: \, H \to H$.) The multiplication law is given by 
\begin{equation}
(h_1,\,g_1)(h_2,\,g_2) = (h_1\, \phi(g_1)h_2,\, g_1g_2).
\end{equation}
This is a rather abstract definition and it takes a bit of effort to prove
that the multiplication law is associative. We will shortly see that in
practice it has a very simple interpretation.

Obviously, for $\phi = \mathrm{id}$ the semidirect product is identical with
the direct product. A useful question for model building is if  
a group can be decomposed into a semidirect product. Actually, a closer
examination of finite groups shows that
semidirect products are ubiquitous! The reason is the following theorem.
\begin{theorem}\label{semidirect}
Let us assume that $H$ is a proper normal subgroup of $S$ and $G$ a subgroup
of $S$ with following properties: 
\begin{enumerate}
\item
$H \cap G = \{ e\}$,
\item
every element $s \in S$ can be written as $s=hg$ with $h \in H$, $g \in
G$.
\end{enumerate}
Then the following holds:
\begin{itemize}
\item
$S \cong H \rtimes_\phi G$ with $\phi(g)h = ghg^{-1}$,
\item
the decomposition $s=hg$ is unique,
\item
$S/H \cong G$.
\end{itemize}
\end{theorem}
\noindent
The proof is straightforward. That the homomorphism $\phi$ has the form given
in the theorem simply follows from the multiplication of two elements of $S$: 
\begin{equation}
s_1 s_2 = (h_1g_1)(h_2g_2) = (h_1 g_1 h_2 g_1^{-1})(g_1g_2).
\end{equation}

\subsection{Symmetries in the Lagrangian versus symmetry groups}
Suppose we have a multiplet of fermion fields $\psi_1, \ldots, \psi_r$ in the
Lagrangian $\mathcal{L}$. Then $\mathcal{L}$ has the form
\begin{equation}
\mathcal{L} = i \sum_{j=1}^r \bar\psi_j \gamma^\mu \partial_\mu \psi_j
  + \cdots
\end{equation}
where the dots indicate the terms beyond the kinetic terms. 
The symmetries of $\mathcal{L}$ are given by transformations
$\psi_j \to A^{(p)}_{jk} \psi_k$ ($p = 1,\ldots,N_\mathrm{gen}$). Since the
kinetic term has to be invariant, it follows that the matrices
$A^{(p)}$ ($p = 1,\ldots,N_\mathrm{gen}$) are unitary.
There are two approaches to symmetries and Lagrangians:
\begin{itemize}
\item[\ding{111}] 
We start with $\mathcal{L}$ and impose symmetries $A^{(p)}$ on $\mathcal{L}$.
Then
the $N_\mathrm{gen}$ matrices $A^{(p)}$ generate a representation of a
symmetry group $G$ from which we can infer the group $G$. 
\item[\ding{111}]
We can also take the opposite point of view. We begin with a group  
$G$ and introduce multiplets of fields which transform according to
representations of $G$. In this way we determine $\mathcal{L}$ from the
symmetry group and the multiplets we introduce.
\end{itemize}

\subsection{Useful theorems for finite groups}
Finite groups, i.e., groups whose number of elements is finite, are very
po\-pular in model building. As expected, 
infinite groups are more complicated than finite ones: They possess 
infinitely many inequivalent irreps and 
non-compact simple Lie groups $G$ possess no finite-dimensional unitary
irreps apart from the trivial ones where every element is mapped onto unity.

Let us for example consider $U(1)$ as the simplest infinite group. We
readily find its irreps: 
$e^{i\alpha} \to e^{in\alpha}$ with $n \in \mathbbm{Z}$. Thus there are
infinitely many. The same applies to the simplest non-abelian group
$O(2)$. Its irreps can be found, for instance, in the appendix of~\cite{o2}.

For finite groups the number of its elements is called order of $G$
and abbreviated by $\ord$. Finite groups have the following properties:
\begin{itemize}
\item
They possess a finite number of inequivalent irreps, 
\item
all irreps are equivalent to unitary irreps,
\item
and all numbers concerning properties of the
group and its irreps are finite as well; this allows to derive extremely
useful relations which are totally lacking in infinite groups.
\end{itemize}

Now we list some of the most important theorems for finite groups:
\begin{theorem}[Lagrange] 
If $H$ is a subgroup of $G$, then $\mbox{ord}\,H$ is a divisor of $\ord$.
\end{theorem}
\noindent
This theorem has a straightforward corollary. 
Defining the order of an element $g$ of $G$ as the smallest number
$r$ such that $g^r = e$, we observe that 
every element $g \in G$ generates a cyclic subgroup 
$\mathbbm{Z}_r \subseteq G$. Therefore, 
the order of every element is a divisor of $\ord$.
\begin{theorem}\label{dim-irreps}
If we denote the irreps of $G$ by $D^{(\alpha)}$, with 
$\mbox{dim}\, D^{(\alpha)} = d_\alpha$ being the dimension of the vector space
on which the irrep acts, and if the
index $\alpha$ numbers all inequivalent irreps, then it follows that 
\begin{equation}
\sum_\alpha d_\alpha^2 = \ord.
\end{equation}
\end{theorem}
\begin{theorem}\label{nc=nirreps}
The number of inequivalent irreps $D^{(\alpha)}$ equals the number of
conjugacy classes of $G$.
\end{theorem}

\subsection{Characters and character tables}
\para{Orthogonality relations for irreps:}
One can define the space of functions on $G$ and endow it with the scalar
product 
\begin{equation}
(f_1 | f_2) = \frac{1}{\ord} \sum_{g \in G} f_1^*(g) f_2(g)
\end{equation}
in order to make it a unitary space. 

Suppose we have an irrep $D^{(\alpha)}$ with dimension $d_\alpha$. Then with
respect to a basis the irrep consists of matrices and we can conceive the
matrix elements $D^{(\alpha)}_{ij}(g)$ as functions on $G$. With Schur's lemma
(not to be confused with theorem 1 (Schur)) it is rather easy to prove the
following theorem~\cite{hamermesh,ramond,ludl}. 
\begin{theorem}
For irreps $D^{(\alpha)}$ and $D^{(\beta)}$ with dimensions $d_\alpha$ and
$d_\beta$, respectively, the orthogonality relations 
\begin{equation}\label{orthogonality}
\sum_{g \in G} D^{(\alpha)}_{ij}(g^{-1}) D^{(\beta)}_{kl}(g) = 
\frac{\ord}{d_\alpha}\, \delta_{\alpha\beta} \delta_{jk} \delta_{il}
\end{equation}
hold.
\end{theorem}
For finite groups we can always assume that the representation matrices
are unitary. In this case 
$D^{(\alpha)}_{ij}(g^{-1}) = ({D^{(\alpha)}}^\dagger)_{ij}(g) = 
(D^{(\alpha)}_{ji}(g))^*$ is valid and equation~(\ref{orthogonality})
can be rewritten as
\begin{equation}
( D^{(\alpha)}_{ji} | D^{(\beta)}_{kl} ) = 
\frac{1}{d_\alpha}\, \delta_{\alpha\beta} \delta_{jk} \delta_{il}.
\end{equation}

\para{The character of a representation:}
For any representation $D$ its character is defined by the function
\begin{equation}
\chi:\;
g \in G  \to  \chi(g) = \mbox{Tr}\, D(g) \in \mathbbm{C}, 
\end{equation}
where Tr denotes the trace.
The character has the property that it is constant on every class $C_k$.

Let us move to the characters of irreps. We denote by $\chi^{(\alpha)}$ the
character of the irrep $D^{(\alpha)}$. These characters have the following
properties: 
\begin{equation}
\chi^{(\alpha)}(e) = d_\alpha, \quad
\sum_{g \in G} \left( \chi^{(\alpha)}(g) \right)^* \chi^{(\beta)}(g)
= \delta_{\alpha\beta}\, \ord.
\end{equation}
The first relation is trivial, the second one follows from
equation~(\ref{orthogonality}). If we denote by 
$c_k$ be the number of elements in class $C_k$ and by $\chi^{(\alpha)}_k$
the value of $\chi^{(\alpha)}$ on $C_k$, then the 
orthogonality relation for the characters of irreps reads
\begin{equation}\label{charo}
\sum_{k=1}^n c_k \left( \chi^{(\alpha)}_k \right)^* \chi^{(\beta)}_k
= \delta_{\alpha\beta}\, \ord,
\end{equation}
where $n$ is the number of classes.

\para{Character tables:}
Since according to theorem~\ref{nc=nirreps} for every group $G$ the number of
classes, $n$, equals the number of inequivalent irreps, 
one can depict a quadratic scheme of numbers $\chi^{(\alpha)}_k$, with 
columns and lines marked by $k$ and $\alpha$, respectively. 
Such a scheme is called character table of the group $G$---see
table~\ref{ct}. Note that this scheme is usually supplemented by two further
lines as shown in table~\ref{ct}, for providing further information on the
group. 
\begin{table}
\begin{center}
\begin{tabular}{|c|cccc|}\hline
$G$        & $C_1$   & $C_2$   & $\cdots$ & $C_n$ \\
(\# $C_k$) & $(c_1)$ & $(c_2)$ & $\cdots$ & $(c_n)$ \\ 
$\oo(C_k)$ & $\nu_1$ & $\nu_2$ & $\cdots$ & $\nu_n$ \\ \hline
\rule{0pt}{5mm}
$D^{(1)}$  & $\chi^{(1)}_1$ & $\chi^{(1)}_2$ & $\cdots$ & $\chi^{(1)}_n$ \\
$D^{(2)}$   & $\chi^{(2)}_1$ & $\chi^{(2)}_2$ & $\cdots$ & $\chi^{(2)}_n$ \\
$\vdots$ & $\vdots$ & $\vdots$ & $\vdots$ & $\vdots$ \\
$D^{(n)}$ & $\chi^{(n)}_1$ & $\chi^{(n)}_2$ & $\cdots$ & $\chi^{(n)}_n$ \\ 
\hline
\end{tabular}
\end{center}
\caption{Schematic description of a character table. In the first line, after
  the name of the group $G$, the classes are listed, below each class $C_k$
  is its number of elements $c_k$, and in the second line below the class the
  order $\nu_k$ of its elements is stated. \label{ct}}
\end{table}
It is customary to set $C_1 = \{ e \}$, thus in the first column the
dimensions $d_\alpha = \chi^{(\alpha)}_1$ of the irreps can be read
off. Furthermore, the usual convention is that $D^{(1)}$ is
the trivial irrep, therefore, $\chi^{(1)}_k = 1$ $\forall k$.
Moreover, the irreps are ordered according to increasing dimensions.

From equation~(\ref{charo}) we know that the line vectors 
\begin{equation} 
\left( \sqrt{\frac{c_1}{\mbox{ord}\, G}} \,
\chi^{(\alpha)}_1, \ldots,
\sqrt{\frac{c_n}{\mbox{ord}\, G}} \, \chi^{(\alpha)}_n
\right),
\end{equation}
form an ON basis of $\mathbbm{C}^n$. Consequently, also 
the column vectors
\begin{equation}
\sqrt{\frac{c_k}{\ord}} \left( \begin{array}{c}
\chi^{(1)}_k \\ \vdots \\ \chi^{(n)}_k 
\end{array} \right) 
\quad (k=1,\ldots,n)
\end{equation}
define an ON basis whose orthonormality conditions can be reformulated as
\begin{equation}\label{charo1}
\sum_{\alpha=1}^n \left( \chi^{(\alpha)}_k \right)^*
\chi^{(\alpha)}_\ell = \frac{\ord}{c_k}\, \delta_{k\ell}.
\end{equation}
Equations~(\ref{charo}) and (\ref{charo1}) are useful for the construction of
a character table.

\para{Reducible representations and character tables:} 
Suppose a representation $D$ of a group $G$ is given. 
Then with its character table it is straightforward to find its decomposition
into irreps because the character of a reducible representation is a sum
\begin{equation}
\chi_D = \sum_{\alpha=1}^n n_\alpha \chi^{(\alpha)}
\end{equation}
where the $n_\alpha$ denote the multiplicities with which the irreps
$D^{(\alpha)}$ occur in $D$. Consequently,
\begin{equation}\label{na}
n_\alpha = ( \chi^{(\alpha)} | \chi_D ).
\end{equation}
This relation is particularly useful for tensor products because the character
of the tensor product $D^{(\alpha)} \otimes D^{(\beta)}$ is given by the
product of the characters of $D^{(\alpha)}$ and $D^{(\beta)}$:
\begin{equation}
\chi^{(\alpha \otimes \beta)}(g) = 
\chi^{(\alpha)}(g) \times \chi^{(\beta)}(g).
\end{equation}

\subsection{The group $S_4$}
Let us examine the symmetric group $S_4$, i.e. the group of permutations of
four objects, in the light of our group-theoretical discussion. We have chosen
$S_4$ for two reasons. First, it is a group which is popular for model
building---see e.g.~\cite{pakvasa} for a very early paper with $S_4$ used in
the quark sector and two recent papers~\cite{lam,GLL} where this group is a
symmetry in the lepton sector. Second, for the symmetric groups $S_n$ there is
a general and simple rule how to find its classes.\footnote{In general, the
  problem of finding the classes of a group can be quite tricky, if its order
  is large.}

The order of $S_n$ is $n!$\hspace{0.7pt}. 
Every element $p \in S_n$ can be written as
\begin{equation}
p = \left( \begin{array}{cccc} 
1   & 2   & \cdots & n \\
p_1 & p_2 & \cdots & p_n 
\end{array} \right). 
\end{equation}
This scheme means that $i$ is mapped to $p_i$ ($i=1,\ldots,n$). One can also
present permutations as cycles. A cycle of length $r$ is a mapping
\begin{equation}
(n_1 \to n_2 \to n_3 \to \cdots \to n_r \to n_1) \equiv 
(n_1n_2n_3 \cdots n_r)
\end{equation}
such that all numbers $n_1$, \ldots, $n_r$ are different. Evidently,  
every permutation is a unique product of cycles which have no common
elements.
For instance, 
\begin{equation}
\left( \begin{array}{cccccc} 
1 & 2 & 3 & 4 & 5 & 6 \\
4 & 6 & 3 & 5 & 1 & 2 
\end{array} \right) = (145)(3)(26).
\end{equation}
Cycles which have no common element commute and 
a cycle which consists of only one element is identical with the unit element
of $S_n$. The classes of $S_n$ are characterized by the cycle
structure~\cite{hamermesh}. 
\begin{theorem}\label{sn-cycles}
The classes of $S_n$ consist of the permutations with the same cycle
structure.
\end{theorem}
Let us apply this to $S_4$. The theorem says that it has five classes
corresponding to the cycle structures 
$e$, $(n_1n_2)$, $(n_1n_2)(n_3n_4)$, $(n_1n_2n_3)$ and $(n_1n_2n_3n_4)$. Its
corresponding classes will be denoted by $C_1, \ldots, C_5$, respectively, in
the following. Thus, $S_4$ has five inequivalent irreps.

There is another useful theorem concerning $S_n$.
\begin{theorem}
$S_n$ has exactly two 1-dimensional irreps: 
$p \to 1$ and $p \to \mbox{sgn}(p)$.
\end{theorem}
\noindent
The sign of a permutation is $+1$ ($-1$), if it can be decomposed into an even
(odd) number of transpositions, i.e. cycles of length $r=2$. 
A cycle of length $r$ is even (odd) if $r$ is odd (even).

Now we can easily find the dimensions of all irreps of $S_4$. We know already
that there are five irreps, with two of them having dimension one. Thus,
according to theorem~\ref{dim-irreps}, we have the equation 
$1^2 + 1^2 + d_3^2 + d_4^2 + d_5^2 = 24$. One can easily check that 
the solution is unique (up to reordering): $d_3 = 2$, $d_4 = d_5 = 3$.

In order to find the remaining three irreps we take advantage of the fact that 
Klein's four-group
\begin{equation}
K = \left\{ e,\, (12)(34),\, (13)(24),\, (14)(23) \right\} \cong \zz \times
\zz
\end{equation}
is a normal, abelian subgroup of $S_4$. That it is an abelian subgroup is
easily checked, that $K$ is also normal follows from theorem~\ref{sn-cycles}. 
We observe that $S_3$ can be conceived as subgroup of $S_4$ if we consider
the permutations of only 2,\,3,\,4. One can check that $K$ and the $S_3$
defined in this way have exactly the properties of $H$ and $G$ of
theorem~\ref{semidirect}. Therefore,
\begin{equation}\label{ks3}
S_4 \cong K \rtimes S_3
\end{equation}
and every element of $S_4$ can uniquely be decomposed into $s = kp$ with 
$k \in K$ and $p \in S_3$.

Taking advantage of equation~(\ref{ks3}), we find the 2-dimensional irrep as
\begin{equation}\label{d2s4}
kp \to D_2(p) \quad \mbox{with} \quad
D_2((234)) = \left( \begin{array}{cc} 
\omega & 0 \\ 0 & \omega^2 
\end{array} \right), \quad 
D_2((34)) = \left( \begin{array}{cc} 
0 & 1 \\ 1 & 0 \end{array} \right).
\end{equation}
Note that $D_2$ is an irrep of $S_3$.
Clearly, $K$, which is represented trivially as we have discussed in
section~\ref{basics}, and the two cycles in equation~(\ref{d2s4}) generate the
full $S_4$, thus we really have found the complete 2-dimensional irrep.

It remains to construct the two 3-dimensional irreps. We only sketch the
procedure. A 3-dimensional representation of $K \cong \zz \times \zz$ 
is given by 
\begin{equation}\label{s4-k}
\renewcommand{\arraystretch}{1.2}
\begin{array}{ccc}
\left( 1 2 \right) \left( 3 4 \right)
&\to& \mathrm{diag} \left( \hphantom{-}1, -1, -1 \right),
\\
\left( 1 3 \right) \left( 2 4 \right)
&\to& \mathrm{diag} \left( -1, \hphantom{-}1, -1 \right),
\\
\left( 1 4 \right) \left( 2 3 \right)
&\to& \mathrm{diag} \left( -1, -1, \hphantom{-}1 \right).
\end{array}
\end{equation}
We denote the representation of $K$ by $A(k)$. Obviously, the mapping 
\begin{equation}\label{s4-t}
\left( 3 4 \right) \to
\left( \begin{array}{ccc}
1 & 0 & 0 \\ 0 & 0 & 1 \\ 0 & 1 & 0
\end{array} \right), 
\quad
\left( 2 4 \right) \to
\left( \begin{array}{ccc}
0 & 0 & 1 \\ 0 & 1 & 0 \\ 1 & 0 & 0
\end{array} \right),
\quad
\left( 2 3 \right) \to 
\left( \begin{array}{ccc}
0 & 1 & 0 \\ 1 & 0 & 0 \\ 0 & 0 & 1
\end{array} \right)
\end{equation}
generates a representation of the $S_3$ which permutes the numbers 
2,\,3,\,4. We denote this representation by $M_3(p)$. It is not difficult to
ckeck that $kp \to A(k)M_3(p)$ is indeed a representation of $S_4$. Obviously,
it is irreducible. The second 3-dimensional irrep is obtained by 
multiplication of the previous one with $\mbox{sgn}(p)$.

Thus we have the following summary of the $S_4$ irreps:
\begin{equation}
\begin{array}{cc}
s = kp \in S_4 \quad \Rightarrow \quad
&
\begin{array}{ll}
\one: & kp \to 1, \\
\one': & kp \to \mbox{sgn}(p), \\
\two: & kp \to D_2(p), \\
\three: & kp \to A(k)M_3(p), \\
\three': & kp \to \mbox{sgn}(p) \,A(k)M_3(p). 
\end{array}
\end{array}
\end{equation}
Note that $\mbox{sgn}(kp) = \mbox{sgn}(p) = \det M_3(p)$.

Having all classes and irreps at our disposal, we can write down 
character table~\ref{ct-s4}. 
\begin{table}
\begin{center}
\begin{tabular}{|c|ccccc|}\hline
$S_4$      & $C_1$ & $C_2$ & $C_3$ & $C_4$ & $C_5$ \\
(\# $C_k$) & (1) & (6) & (3) & (8) & (6) \\
$\oo(C_k)$ & 1 & 2 & 2 & 3 & 4 \\\hline
\rule{0pt}{4mm}
$\one$ & 1 & 1 & 1 & 1 & 1 \\
$\one'$ & 1 & $-1$ & $1$ & 1 & $-1$ \\
$\two$ & 2 & 0 & 2 & $-1$ & 0 \\
$\three$ & 3 & 1 & $-1$ & 0 & $-1$ \\
$\three'$  & 3 & $-1$ & $-1$ & 0 & $1$ \\\hline
\end{tabular}
\end{center}
\caption{Character table of $S_4$. \label{ct-s4}}
\end{table}
As an application we compute the decomposition of $\three \otimes \three$ into
irreps. The character of $\three \otimes \three$ is given by the square of the
line labeled by $\three$ in table~\ref{ct-s4}:
\begin{equation}\label{33}
\chi^{\three \otimes \three} = \left[\, 9 \; 1 \; 1 \; 0 \; 1 \,\right].
\end{equation}
With equation~(\ref{na}) the multiplicities of the irreps in 
$\three \otimes \three$ are computed. The whole information for this
computation is contained in the character table:
\begin{align}
n_\one &= \frac{1}{24} \left( 
1 \times 1 \times 9 + 6 \times 1 \times 1 + 3 \times 1 \times 1 + 
8 \times 1 \times 0 + 6 \times 1 \times 1 
\right) = 1, \nonumber \\
n_{\one'} &= \frac{1}{24} \left( 
1 \times 1 \times 9 - 6 \times 1 \times 1 + 3 \times 1 \times 1 + 
8 \times 1 \times 0 - 6 \times 1 \times 1 
\right) = 0, \nonumber \\
n_\two &= \frac{1}{24} \left( 
1 \times 2 \times 9 + 6 \times 0 \times 1 + 3 \times 2 \times 1 - 
8 \times 1 \times 0 + 6 \times 0 \times 1 
\right) = 1, \nonumber \\
n_\three &= \frac{1}{24} \left( 
1 \times 3 \times 9 + 6 \times 1 \times 1 - 3 \times 1 \times 1 + 
8 \times 0 \times 0 - 6 \times 1 \times 1 
\right) = 1, \nonumber \\
n_{\three'} &= \frac{1}{24} \left( 
1 \times 3 \times 9 - 6 \times 1 \times 1 - 3 \times 1 \times 1 + 
8 \times 0 \times 0 + 6 \times 1 \times 1 
\right) = 1. \nonumber 
\end{align}
All products of three numbers in this computation are given by 
\[
c_k \times \chi^{(\alpha)}_k \times \chi^{\three \otimes \three}_k.
\]
Thus the result of the decomposition is
\begin{equation}
\three \otimes \three = \one \oplus \two \oplus \three \oplus \three'.
\end{equation}
With some experience it is not difficult to guess the Clebsch--Gordan
coef\-fi\-cients---for their definition see e.g. \cite{hamermesh,ramond}. 
Denoting the cartesian basis vectors in $\three$ by 
$e_j$ ($j=1,2,3$) and defining $\omega = e^{2\pi i/3}$ we find 
\begin{equation}\label{3x3-s4}
\renewcommand{\arraystretch}{1.2}
\begin{array}{rc}
\one: & \frac{1}{\sqrt{3}} \left( 
e_1 \otimes e_1 + e_2 \otimes e_2 + e_3 \otimes e_3 \right), \\
\two: & \left\{ \begin{array}{c}
\frac{1}{\sqrt{3}} \left( e_1 \otimes e_1 + \omega^2 e_2 \otimes e_2 + \omega 
e_3 \otimes e_3 \right), \\
\frac{1}{\sqrt{3}} \left( e_1 \otimes e_1 + \omega e_2 \otimes e_2 + \omega^2 
e_3 \otimes e_3 \right),
\end{array} \right.
\\
\three: & \left\{ \begin{array}{c}
\frac{1}{\sqrt{2}} \left( e_2 \otimes e_3 + e_3 \otimes e_2 \right), \\ 
\frac{1}{\sqrt{2}} \left( e_3 \otimes e_1 + e_1 \otimes e_3 \right), \\ 
\frac{1}{\sqrt{2}} \left( e_1 \otimes e_2 + e_2 \otimes e_1 \right), 
\end{array} \right.
\\
\three': & \left\{ \begin{array}{c}
\frac{1}{\sqrt{2}} \left( e_2 \otimes e_3 - e_3 \otimes e_2 \right), \\ 
\frac{1}{\sqrt{2}} \left( e_3 \otimes e_1 - e_1 \otimes e_3 \right), \\ 
\frac{1}{\sqrt{2}} \left( e_1 \otimes e_2 - e_2 \otimes e_1 \right). 
\end{array} \right.
\end{array}
\end{equation}

\subsection{The group $A_4$}
After the seminal paper by Ma and Rajasekaran~\cite{ma}, this group has become
the most popular one in the context of neutrino masses and lepton mixing. 
We can only list a few early papers here in~\cite{A4,he}, refer the reader to
the review~\cite{review} and to citations in recent $A_4$ papers to get an
impression of the bustling activities with respect to model building with
$A_4$. It is worth noting that this group has already been used much earlier
in the quark sector~\cite{wyler}.

The group $A_4$ consists of all even permutations of $S_4$. Therefore, its
structure is 
\begin{equation}\label{a4-sd}
A_4 \cong K \rtimes \mathbbm{Z}_3.
\end{equation}
Theorem~\ref{sn-cycles} cannot be applied to find the classes, it is however
clear that the classes of $A_4$ must be subsets of the classes of $S_4$ which
consist of even permutations. In this way we obtain
\begin{equation}
\begin{array}{ccl}
C_1 &=& \{ e \}, \\
C_2 &=& \{(12)(34),\, (13)(24),\, (14)(23) \},  \\
C_3 &=& \{ (132),\, (124),\, (234),\, (143) \}, \\
C_4 &=& \{ (123),\, (142),\, (243),\, (134) \}.
\end{array}
\end{equation}
Thus we know that $A_4$ has four inequivalent irreps.
Equation~(\ref{a4-sd}) tells us that there are three 1-dimensional irreps
stemming from the $\mathbbm{Z}_3$, which map $K$ onto 1:
\begin{equation}
\one: \, (243) \to 1, \quad 
\one': \, (243) \to \omega^2, \quad
\one'': \, (243) \to \omega.
\end{equation}
Since $A_4$ has 12 elements, the remaining irrep must have dimension three. 
Equations~(\ref{s4-k}) and (\ref{s4-t}) for $S_4$ allow to determine this
irrep: 
\begin{equation}\label{AE}
(12)(34) \to A \equiv \left( \begin{array}{ccc}
1 & 0 & 0 \\ 0 & -1 & 0 \\ 0 & 0 & -1
\end{array} \right), \quad
(243) \to E = \left( \begin{array}{ccc}
0 & 1 & 0 \\ 0 & 0 & 1 \\ 1 & 0 & 0
\end{array} \right).
\end{equation}
For the second relation we have exploited the relation $(243) = (23)(24)$.
An alternative definition of $A_4$ is given by this irrep 
because the $\three$ is faithful. In this way, $A_4$ can be considered as a
finite subgroup of $SU(3)$ with generators $A$ and $E$.

Having constructed all irreps we can write down the character table of
$A_4$---see table~\ref{ct-a4}.
\begin{table}
\begin{center}
\begin{tabular}{|c|cccc|}
\hline 
$A_4$ & $C_1$ &  $C_2$ & $C_3$ & $C_4$ \\
(\# $C_k$) & (1) & (3) & (4) & (4) \\
$\oo (C_k)$ & 1 & 2 & 3 & 3 \\
\hline 
$\mathbf{1}$ & 1 & 1 & 1 & 1\\
$\mathbf{1}'$ & 1 & 1 & $\omega$ & $\omega^2$ \\
$\mathbf{1}''$ & 1 & 1 & $\omega^2$ & $\omega$ \\
$\mathbf{3}$ & 3 & $-1$ & 0 & 0 \\
\hline 
\end{tabular}
\end{center}
\caption{Character table of $A_4$. \label{ct-a4}}
\end{table}
As an example for its usage one can, for instance, compute 
\begin{equation}
\three \otimes \three = \one \oplus \one' \oplus \one'' \oplus \three
\oplus \three.
\end{equation}
The Clebsch--Gordan decomposition of this tensor product is given by
\begin{equation} 
\renewcommand{\arraystretch}{1.2}
\begin{array}{rl}
\one: & \frac{1}{\sqrt{3}} \left( 
e_1 \otimes e_1 + e_2 \otimes e_2 + e_3 \otimes e_3 \right), \\
\one': & \frac{1}{\sqrt{3}} \left( 
e_1 \otimes e_1 + \omega^2 e_2 \otimes e_2 + \omega 
e_3 \otimes e_3 \right), \\
\one'': & \frac{1}{\sqrt{3}} \left( 
e_1 \otimes e_1 + \omega e_2 \otimes e_2 + \omega^2 
e_3 \otimes e_3 \right), \\
\three: & 
e_2 \otimes e_3,\, e_3 \otimes e_1,\, e_1 \otimes e_2, \\ 
\three: &
e_3 \otimes e_2,\, e_1 \otimes e_3,\, e_2 \otimes e_1.
\end{array}
\end{equation}
For the two 3-dimensional irreps one could equivalently use the symmetric and
antisymmetric combinations of $e_j \otimes e_k$, or any other weighted,
orthogonal combination. Note that
for $S_4$ one does not have this freedom, one must use the symmetric
combination for the $\three$ and the antisymmetric combination for the
$\three'$---see equation~(\ref{3x3-s4}). 
In the case of $A_4$ this freedom comes about because the $\three'$
becomes identical with the $\three$ due to the absence of transpositions.

\section{Models of neutrino masses and lepton mixing}
\subsection{Lagrangians and horizontal symmetries}

We begin with some remarks. The notion ``horizontal symmetry'' is used
synonymously with ``family symmetry'':
\begin{itemize}
\item[\ding{60}]
We assume that any model we have in mind is an extension of the Standard
Model. Therefore, the full symmetry group of the Lagrangian $\mathcal{L}$ is 
$G_\mathrm{gauge} \times G_\mathrm{family}$. 
($G_\mathrm{family}$ could also be gauged, but we do not consider this
possibility here.) 
\item[\ding{60}]
Kinetic and gauge terms in the Lagrangian are automatically invariant under
$G_\mathrm{family}$. 
\item[\ding{60}]
Therefore, the effect of $G_\mathrm{family}$ is felt in the Yukawa Lagrangian
and the scalar potential.
\item[\ding{60}]
The Yukawa couplings are connected with the Clebsch--Gordan coefficients of 
the tensor products of the fermion representations, such that for every irrep
of scalar fields there is a free Yukawa coupling constant.
\item[\ding{60}]
The mass matrices contain, in addition, the vacuum expectation values (VEVs)
which are determined by the minimum of the scalar potential.
\item[\ding{60}]
With several VEVs one has the problem of \emph{vacuum alignment}. The meaning
of this notion is that only specific VEV relations lead to mass
matrices which give the desired mixing angles and, sometimes in addition,
predictions for the neutrino mass spectrum.
\item[\ding{60}]
With family symmetries one has almost necessarily a proliferation of the scalar
sector and, in most cases, also  additional fermion fields. Thus there is a
tension between the introduction of new fields and, as a consequence, unknown
constants, which are necessary to realize the symmetry, and the attempted 
predictions for masses and mixings.
\end{itemize}

Let us discuss the relation between Clebsch--Gordan coefficients and  Yukawa
couplings in more detail. Suppose we have a 
tensor product $D \otimes D' = D_S \oplus \cdots$ with irreps 
$D$, $D'$ and $D_S$. We choose the bases $D: \, \{ e_\alpha \}$ and 
$D': \, \{ f_\alpha \}$. Then the basis for irrep 
$D_S$ has the form 
$\{ b_i = \Gamma_{i\alpha\beta} e_\alpha \otimes f_\beta \}$.
With the transformations 
\begin{equation}
e_\alpha \to D_{\gamma\alpha} e_\gamma, \quad 
f_\beta \to D_{\delta\beta} f_\delta, \quad 
b_i \to (D_S)_{ji} b_j
\end{equation}
the conditions on the Clebsch--Gordan  coefficient matrices $\Gamma_i$ are
obtained as 
\begin{equation}\label{cond}
\Gamma_i = \left( D^\dagger \Gamma_j {D'}^* \right) (D_S)_{ji}.
\end{equation}
Now we consider generic Yukawa couplings in the Majorana form 
\begin{equation}
\mathcal{L}_Y = y\, \psi_\alpha^T C^{-1} \gamma_{i\alpha\beta} S_i\, 
\psi'_\beta + \mbox{H.c.},
\end{equation}
where $\psi$ and $\psi'$ transform according to $D$ and $D'$,
respectively. Comparing with equation~(\ref{cond}) we find that
\begin{equation}
\psi \to D \psi, \quad D' \to D' \psi' \quad \Rightarrow \quad
S \to D_S^* S, \quad \gamma_i = \Gamma_i^*.
\end{equation}
I.e., the scalar fields transform with the irrep complex conjugate to $D_S$ and
the Yukawa couplings are partially determined by the complex conjugate 
Clebsch--Gordan coefficient matrix, as announced above.

If we have three fermion families then the fermion multiplets constitute
3-dimensional representations of the horizontal group $G$. We can distinguish
three cases:
\begin{enumerate}
\renewcommand{\theenumi}{\roman{enumi}}
\item 
Abelian case: Only 1-dimensional irreps are present.
\item Non-Abelian case: 2-dimensional irreps occur, but no 3-dimensional one.
\item Non-Abelian case: 3-dimensional irreps occur.
\end{enumerate}
An Abelian group $G$ is 
synonymous with ``texture zeros'', i.e., a Yukawa coupling is either present
and undetermined or it is zero, but there are no relations between different
Yukawa couplings. Relations among observables in the mass spectrum and
mixing have their origin solely in these zeros. 
It has been shown~\cite{low} that by Abelian symmetries the only extremal
mixing angle which can be enforced is $\theta_{13} = 0^\circ$.
It is possible to enforce texture zeros in arbitrary entries of the fermion
mass matrices by means of Abelian symmetries and an extended scalar
sector~\cite{joshipura}. 

In the second case it is possible to enforce 
$\theta_{13} = 0^\circ$ \emph{and} $\theta_{23} = 45^\circ$. For tri-bimaximal
mixing one needs 3-dimensional irreps. In the next subsection we will discuss
one such model based on $A_4$.

\subsection{A type~I seesaw model based on $A_4$}

As a prototype for a renormalizable $A_4$ model we discuss the model
of~\cite{he}. It is based on the following $A_4$ multiplets:
\begin{equation}
\begin{array}{rl}
\mbox{fermion fields:} &
\ell_R \in \one \oplus \one' \oplus \one'', \quad
D_L \in \three, \quad \nu_R \in \three, \\
\mbox{scalar fields:} &
\phi \in \three, \quad \phi_0 \in \one, \quad
\chi \in \three.
\end{array}
\end{equation}
In this list the $D_L$ are the usual leptonic left-handed gauge doublets, the
$\ell_R$ are the right-handed charged gauge singlets and the $\nu_R$ are the
right-handed neutrino singlets. There are four Higgs doublets $\phi$ and
$\phi_0$ with hypercharge $+1$ and three real gauge singlets $\chi$. 

With the discussion in the previous section it is straightforward to derive
the Lagrangian
\begin{eqnarray}
\mathcal{L} &=& \cdots -\left[
h_1 \left( \bar D_{1L} \phi_1 + \bar D_{2L} \phi_2 + \bar D_{3L}
\phi_3 \right) \ell_{1R} \right. \nonumber \\ 
&& \hphantom{xxx} + 
h_2 \left( \bar D_{1L} \phi_1 + \omega^2 \bar D_{2L} \phi_2 + 
\omega \bar D_{3L} \phi_3 \right) \ell_{2R} \nonumber \\ 
&&  \hphantom{xxx} + 
h_3 \left( \bar D_{1L} \phi_1 + \omega \bar D_{2L} \phi_2 + 
\omega^2 \bar D_{3L} \phi_3 \right) \ell_{3R} 
\label{y1}\\ 
&& \hphantom{xxx} + 
h_0 \left( \bar D_{1L} \nu_{1R} + \bar D_{2L} \nu_{2R} + \bar D_{3L}
\nu_{3R} \right) \tilde \phi_0 + \mbox{H.c.} \big]
\label{y0}
\\ && +
\frac{1}{2} \left[ M \left( \nu_{1R}^T C^{-1} \nu_{1R} +  
\nu_{2R}^T C^{-1} \nu_{2R} +  \nu_{3R}^T C^{-1} \nu_{3R} \right) + 
\mbox{H.c.} \right]
\label{mass}
\\ && +
\frac{1}{2} \left[ h_\chi 
\left( \chi_1 \left( \nu_{2R}^T C^{-1} \nu_{3R} 
+  \nu_{3R}^T C^{-1} \nu_{2R} \right) \right. \right. 
\nonumber \\ 
&& \hphantom{xxxii} +
\chi_2 \left( \nu_{3R}^T C^{-1} \nu_{1R} + \nu_{1R}^T C^{-1} \nu_{3R}
\right) 
\nonumber 
\\ && \left. \left. \hphantom{xxxii} + 
\chi_3 \left( \nu_{1R}^T C^{-1} \nu_{2R} + \nu_{2R}^T C^{-1} \nu_{1R}
\right) \right) + 
\mbox{H.c.} \right],
\label{ychi}
\end{eqnarray}
where we have confined ourselves to the Yukawa interactions and mass
terms. The dots indicate the kinetic terms, the gauge interactions and the
scalar potential. 

Through spontaneous symmetry breaking with VEVs $v_j$, $w_j$ ($j=1,2,3$) and
$v_0$ of the Higgs doublets and scalar singlets, respectively, 
the Lagrangian leads to the mass terms
\begin{equation}
-\bar\ell_L M_\ell \ell_R - \bar\nu_L M_D \nu_R + 
\frac{1}{2} \nu_R^T C^{-1} M_R \nu_R + \mbox{H.c.}
\end{equation}
While $M_D = h_0 v_0^* \mathbbm{1}$ is simply proportional to the unit matrix,
the other two mass matrices are given by
\begin{equation}\label{MM}
M_\ell = \left(
\begin{array}{ccc}
h_1 v_1 & h_2 v_1          & h_3 v_1 \\
h_1 v_2 & h_2 v_2 \omega^2 & h_3 v_2 \omega \\
h_1 v_3 & h_2 v_3 \omega   & h_3 v_3 \omega^2 
\end{array} \right)
\;\, \mbox{and} \;\,
M_R = \left( \begin{array}{ccc}
M & h_\chi w_3 & h_\chi w_2 \\
h_\chi w_3 & M & h_\chi w_1 \\
h_\chi w_2 & h_\chi w_1 & M
\end{array} \right).
\end{equation}

With general VEVs one cannot obtain tri-bimaximal mixing. It is well known
that the vacuum alignment
\begin{equation}\label{va}
v_1 = v_2 = v_3 \equiv v, \quad w_1 = w_3 = 0, \quad 
h_\chi w_2 \equiv M'
\end{equation}
is needed, which gives the mass matrices
\begin{equation}\label{mlmr}
M_\ell = \sqrt{3} v\, U_\omega^\dagger \left( \begin{array}{ccc}
h_1 & 0 & 0 \\ 0 & h_2 & 0 \\ 0 & 0 & h_3
\end{array} \right),
\quad
M_R = \left( \begin{array}{ccc}
M & 0 & M' \\
0 & M & 0 \\
M' & 0 & M
\end{array} \right).
\end{equation}
The matrix $U_\omega$ can be read off from $M_\ell$ in equation~(\ref{MM}).
We denote by $U_\nu$ the matrix which diagonalizes $M_R$ of
equation~(\ref{mlmr}). These two unitary matrices are then obtained as
\begin{equation} 
U_\omega = 
\frac{1}{\sqrt{3}} \left( \begin{array}{ccc} 
1 & 1 & 1 \\ 1 & \omega & \omega^2 \\ 1 & \omega^2 & \omega
\end{array} \right),
\quad
U_\nu = \left( \begin{array}{ccc}
1/\sqrt{2} & 0 & -1/\sqrt{2} \\
0 & 1 & 0 \\
1/\sqrt{2} & 0 & 1/\sqrt{2} 
\end{array} \right).
\end{equation}
Therefore, up to diagonal phase matrices on the left and right-hand side, we
arrive at the lepton mixing matrix 
\begin{equation}\label{UU}
U = U_\omega U_\nu = 
\mbox{diag}\,(1,\omega,\omega^2)\, U_\mathrm{HPS} \,
\mbox{diag}\,(1,1,-i).
\end{equation}
Thus the $A_4$ symmetry of the Lagrangian together with suitable vacuum
alignment leads to tri-bimaximal mixing. The charged-lepton masses are
reproduced by choosing the Yukawa couplings appropriately: 
$m_\alpha = \sqrt{3} |v h_\alpha |$ ($\alpha = e,\mu,\tau$).

We conclude the discussion of the model of~\cite{he} with a few comments.
As shown above, vacuum alignment is a very important ingredient for achieving
tri-bimaximal mixing. Of course, the symmetry group also restricts the scalar
potential and is essential for allowing the required vacuum structure to be a
minimum of the scalar potential. Nevertheless, vacuum alignment is usually a
tricky problem. In the model we have discussed it was necessary to break $A_4$
down to $\mathbbm{Z}_3$ generated by the matrix $E$---see
equation~(\ref{AE})---in the charged lepton 
sector, while in the neutrino sector the VEVs of the scalars $\chi_k$ break
$A_4$ to a $\zz$ generated by $\mbox{diag}\,(-1,\,1,-1)$. This is quite generic
for $A_4$ models with tri-bimaximal mixing and leads to $M_D \propto
\mathbbm{1}$ and the structure of $M_\ell$ and $M_R$ of
equation~(\ref{mlmr}). It was shown
in~\cite{he} that the vacuum alignment~(\ref{va}) is possible if the scalar
potential is CP-conserving. 
Since in this model 
$\ell_R$ is not in the same $A_4$ multiplet as $D_L$ and $\nu_R$, it
cannot be embedded in a Grand Unified Theory. Note, however, that it is
possible to put both $D_L$ and $\ell_R$ into a $\three$ and to use the type~II
seesaw mechanism with scalar gauge triplets--see for instance~\cite{triplet}, 
a scenario which can at least in principle be extended to a Grand Unified
Theory.

\section{Conclusions}

A large part of this lecture dealt with the theory of finite groups, the
other part with the application of group theory to Lagrangians
for the purpose of ``explaining'' mass and mixing patterns found experimentally
or to make predictions in this context. Let us finish with remarks on the
second part.

We have tried to demonstrate that 
symmetries based on finite groups \emph{could} be a way to tackle the mass
and mixing problem. However, all 
models for lepton mixing (and neutrino masses)
require complicated and contrived extensions of the Standard Model. 
Such models are in most cases incompatible with Grand Unification,
need vacuum alignment, employ SUSY and non-renormalizable terms, etc.
Here we have confined ourselves to the re\-latively simple renormalizable
model of~\cite{he} as a showcase, which manages without SUSY. 
As for tri-bimaximal mixing~(\ref{HPS}), for the time being it 
is compatible with all experimental results. 
However, it could turn out that $s_{13}^2 \sim 0.01$~\cite{results}.
In that case, ideas alternative to tri-bimaximal mixing would be in demand. 
Or one assumes that tri-bimaximal mixing holds at a high (seesaw) scale 
and, by the renormalization group evolution of the mixing
angles from the high scale down to the electroweak scale, $s_{13}^2$ evolves
sufficiently away from zero; this is possible with a degenerate  
neutrino mass spectrum.

\para{Acknowledgments:} The author thanks the organizers for their kind
hospitality and the pleasant atmosphere at the school. Furthermore, he is very
grateful to Patrick Ludl for a careful reading of the manuscript.

\end{document}